\documentclass[12pt]{article}
\usepackage[body={18cm, 23cm, centered}]{geometry}
\usepackage{amsmath,amssymb,amsthm,amscd,cite,comment,amsfonts,indentfirst,color,setspace,bbold,arydshln}
\usepackage{mathtext,marvosym,textcomp}
\usepackage[makeroom]{cancel}
\usepackage{mathtools}

\usepackage[debug,pageanchor=false]{hyperref}
\hypersetup{colorlinks=true,linktocpage,breaklinks,
	urlcolor=blue,
	linkcolor=red,
	citecolor=blue
}

\usepackage[vcentermath]{youngtab}

\usepackage[mathmode,centertableaux]{ytableau}
\usepackage{tikz}
\usetikzlibrary{arrows}
\usetikzlibrary{shapes.misc}
\usepackage{tikz-cd}

\tikzset{cross/.style={cross out, draw=black, minimum size=2*(#1-\pgflinewidth), inner sep=0pt, outer sep=0pt},
	cross/.default={5pt}}
\usepackage[ normalem]{ulem}
\usepackage{tocloft}

\numberwithin{equation}{section}

\selectfont

\def\a{\alpha} 
\def\b{\beta} 
 
\def\d{\delta} 
\def\e{\epsilon}

\def\h{\eta}

\def\m{\mu}
\def\n{\nu}

\def\q{\theta}
\def\s{\sigma} 
\def\t{\tau}  
\def\f{\phi}
 
\def\w{\omega}

\def\D{\Delta}

\def\S{\Sigma}

\def\fr{\frac}  \def\dt{\partial}

\def\mc{\mathcal}

\def\mF{\mathcal{F}}

\def\mH{\mathcal{H}}

\def\mH{\mathcal{H}}

\def\mR{\mathcal{R}}
\def\mM{\mathcal{M}}
\def\tf{\tilde{f}}

\def\tx{\tilde{x}}
\def\ty{\tilde{y}}

\def\tT{\tilde{T}}

\def\M{{\mathcal{M}}}

\def\XX{\mathbb{X}}

\def\ZZ{\mathbb{Z}}
\def\TT{\mathbb{T}}

\def\Tr{\mathrm{Tr}}

\def\gg{\mathfrak{g}}
\def\tgg{\tilde{\mathfrak{g}}}
\def\sl{\mathfrak{sl}}
\def\gl{\mathfrak{gl}}
\def\so{\mathfrak{so}}

\def\ha{\hat{\alpha}} 
\def\hb{\hat{\beta}}

\def\hgr{\hat{g}}
\def\hhr{\hat{h}}
\def\tgr{\tilde{g}} 
\def\thr{\tilde{h}} 

\def\bgreen{green!70!black}

\def\bas{{\mathrm{bas}}}

\begin{document}
\renewcommand{\refname}{\begin{center}References\end{center}}
	
\begin{titlepage}
		
	\vfill
	\begin{flushright}

	\end{flushright}
		
	\vfill
	
	\begin{center}
		\baselineskip=16pt
		{\Large \bf 
			On non-abelian U-duality of 11D backgrounds
		}
		\vskip 1cm
			Edvard T. Musaev
		\vskip .3cm
		\begin{small}
			{\it 
			    Moscow Institute of Physics and Technology,  Institutskii per. 9, Dolgoprudny, 141700, Russia\\
			    Kazan Federal University, Institute of Physics,  Kremlevskaya 16a, Kazan, 420111, Russia
			}
		\end{small}
	\end{center}
		
	\vfill 
	\begin{center} 
		\textbf{Abstract}
	\end{center} 
	\begin{quote}
         In this letter we generalised the  procedure of non-abelian T-duality based on a B-shift and a sequence of formal abelian T-dualities in non-isometric directions to 11-dimensional backgrounds. This consists of a C-shift followed by either a formal U-duality transformation or taking a IIB section. We investigate restrictions and applicability of the procedure and find that it can provide supergravity solutions for the SL(5) exceptional Drinfeld algebra only when a spectator field is present, which is consistent with examples known in the literature.
	\end{quote} 
	\vfill
	\setcounter{footnote}{0}
\end{titlepage}
	
\clearpage
\setcounter{page}{2}
	
\tableofcontents

\section{Introduction}

String theory is known to respect a rich set of various symmetries, among which those that transform target space-time keeping physics the same are of special interest. The most known example of such duality symmetries is the perturbative T-duality symmetry of Type II string theory, that acts along toroidal directions of target-space according to the so-called Buscher rules \cite{Buscher:1987qj,Buscher:1987sk}. The procedure for recovering background fields transformations from the string partition function is well known. One starts with the string  partition function defined by the action $S_0[\q]$ symmetric under global $\q \to \q+\a$ with $\q$ corresponding to a circular direction. The symmetry is then gauged by introducing a 1-form field $d\q \to D\q=d\q+A$ and the corresponding Lagrange term $\tilde{\q} F$ with $F=d A$ to keep the 1-form pure gauge. The resulting partition function defined by the action $S_1[\q,A,\tilde{\q}]$ can then be reduced to the initial one, integrating out $\tilde{\q}$, that sets $A=d\a$. Alternatively, one integrates out the 1-form field $A$ obtaining a string action $S_2[\tilde{\q}]$ defined on a different background related to the initial one by Buscher rules. The scalar field $\q(\s,\t)$ gets replaced by the field $\tilde{\q}(\s,\t)$ representing dual string coordinates corresponding to winding modes \cite{Fradkin:1984ai,Tseytlin:1990nb}. Transformation of the dilaton ensures that measure in the partition function is invariant at one loop. One can be more general and consider backgrounds of the form $M\times \TT^d$ in which case T-duality group will be O$(d,d;\ZZ)$.

A natural question is whether one may consider backgrounds with isometries represented by more involved groups than abelian U$(1)^d$, say a sphere or a non-abelian group manifold. The answer is positive and the corresponding dualisation procedure has been considered in \cite{delaOssa:1992vc}. Essentially non-abelian T-duality of string partition function goes along the same lines as the abelian one. The difference comes from more involved definition of the field strength $F=dA+[A,A]$, that is now an element of the corresponding algebra and hence the Lagrange term reads $\Tr[\tilde{\q}F]$. Hence, one dualises the whole set of group coordinates basically replacing left-invariant 1-forms $\s^a$ by dual forms $d\tilde{\q}_a$. The original procedure for NS-NS fields has been complemented by transformation rules for RR fluxes in \cite{Sfetsos:2010uq,Lozano:2011kb}. Explicit canonical formulation of non-abelian T-duality for principal sigma-model has been provided in \cite{Lozano:1995jx}. Additionally, the work \cite{Lozano:2011kb} provided a procedure of non-abelian T-dualisation for coset space geometries $G/H$ based on fixing gauge degrees of freedom  corresponding to action of the subgroup $H$. In contrast to abelian T-duality its non-abelian generalisation does not preserve isometries of the original background (in the usual sense) and hence has many in common with deformations of supergravity backgrounds. In particular, NATD techniques have been widely used to generate new supergravity backgrounds interesting from the point of view of holography, and in \cite{Hoare:2016wca} some explicit examples of such relation have been provided.

Breaking of the initial background isometries by a non-abelian T-duality transformation is in severe contrast with mechanics of the standard abelian T-duality transformations, where preservation of isometries allows to perform T-duality twice making it an involutive symmetry. For a way out of this problem, one looks at Noether currents of the two-dimen\-sional string sigma-model and their Bianchi identities. Starting with sigma-model on a background with isometry algebra defined by structure constants $f_{ab}{}^c$ one is able to construct conserved Noether currents $J_a$, that satisfy
\begin{equation}
    dJ_a=0.
\end{equation}
Non-abelian T-dualising along the isometry directions one ends up with sigma-model on a background with no initial isometries, which however still allows to define Noether currents $J_a$, that satisfy \cite{Klimcik:1995dy}
\begin{equation}
    dJ_a=\tilde{f}_a{}^{bc}J_b \wedge J_c.
\end{equation}
Here the algebras $\mathfrak{g}$ and $\tilde{\mathfrak{g}}$ defined by the structure constants $f_{ab}{}^c$ and $\tilde{f}_a{}^{bc}$ form the so-called Drinfeld double $\mc{D}$. This is defined as a Manin triple $(\mc{D},\gg,\tgg)$ with the non-degenerate form given by the $O(d,d)$ invariant metric $\h$. Such algebraic construction allows to reverse the NATD transformation applying a Poisson-Lie T-duality transformation, that basically means solving consistency constraints for the Drinfeld double and constructing a background with such isometries (dressing the generalised vielbein in DFT terms). More details on Poisson-Lie T-duality and NATD can be found in the original works \cite{Klimcik:1995ux,VonUnge:2002xjf} and in review papers \cite{Petr:2010nzh,Sfetsos:2011jw,Thompson:2019ipl}. For developments from the generalised geometry side one refers to \cite{Hassler:2017yza,Alfonsi:2019ggg,Catal-Ozer:2019hxw,Demulder:2019bha}. Explicit examples of backgrounds resulting from PLTD and/or NATD can be found in \cite{Hlavaty:2020afj,Lozano:2011kb,Hong:2018tlp,Hlavaty:2019pze,Hlavaty:2015hla, Eghbali:2013bda}. Representation of Yang-Baxter bi-vector deformations as a B-shift followed by an NATD transformation has been considered in \cite{Borsato:2018idb}.

To some extent the above constructions generalise to M-theory in the sense of membrane dynamics and 11-dimensio\-nal supergravity. From the membrane point of view non-abelian U-duality have been addressed in \cite{Sakatani:2020iad}, where in particular an analogue of Bianchi identities for currents of 2-dimensional sigma-model have been derived and implemen\-ted into the SL(5) exceptional field theory. The notion of Drinfeld double (Manin triple) have been generalised to the so-called exceptional Drinfeld algebra in the series of works \cite{Sakatani:2019zrs,Malek:2019xrf}, which however does not carry the structure of a bi-algebra. Instead, the algebra $\tgg$ dual to the isometry algebra $\gg$ is defined via tri-algebra structure constants $\tilde{f}_a{}^{bcd}$, that is in consistency with the current algebra of \cite{Sakatani:2020iad}. Finally, certain explicit results for  non-abelian U-dualised backgrounds and their relation to non-abelian T-duality have been presented recently in \cite{Blair:2020ndg}.

This letter considers a generalisation of the approach of \cite{Borsato:2018idb} to non-abelian T-duality in the formalism of exceptional field theory. In \cite{Borsato:2018idb} explicit Buscher rules for non-abelian T-duality transformation have been provided written in terms of undressed fields, that can be represented as O$(d,d)$ transformations of the corresponding generalised metric of double field theory \cite{Sakatani:2019jgu,Catal-Ozer:2019hxw}. Dependence on parameters $\tilde{x}_a$ enters in the final expression that finally get interpreted as dual coordinates. Given the embedding into DFT the procedure can be generalised to M-theory backgrounds in terms of exceptional field theory generalised metrics and dual coordinates $\tilde{x}_{ab}$ corresponding to winding modes of membranes.

The text is structured as follows. In Section \ref{sec:natd} we review the NATD procedure as an O(D,D) rotation for group manifolds. As an explicit example Bianchi II space-time with vanishing dilaton is considered. In Section \ref{sec:naud} we generalise the approach to non-abelian U-duality transformations of 11-dimensional backgrounds. In Section \ref{sec:alg} we analyse the suggested procedure for ExFT's based on U-duality groups SL(5) and SO(5,5) and derive conditions upon which a solution of 11-dimensional supergravity can be generated.

\section{Non-abelian T-duality}
\label{sec:natd}

\subsection{Sigma-model perspective}

Non-abelian T-duality transformations generalise standard T-duality Buscher rules and can be written in a very similar form \cite{Borsato:2018idb}. The case of our interest here is backgrounds of the form $M\times G$ where G is a group manifolds, however the sigma-model procedure can be generalised to coset spaces. To set up the notations we briefly discuss the procedure of \cite{Borsato:2018idb} here. One starts with the sigma model action of the form
\begin{equation}
    S=T \int_\S \Big(\fr12 E^{\ha}\wedge * E^{\hb}\h_{\ha\hb}+B\Big),
\end{equation}
where the vielbein 1-form $E^{\ha}$ is defined as usual as
\begin{equation}
    \begin{aligned}
    E^{\ha}&=(g^{-1}dg)^a E_a{}^{\ha}+dx^\m E_\m{}^{\ha}, && g\in G,\\
    g^{-1}dg&= \s_m{}^a dy^m T_a.
    \end{aligned}
\end{equation}
Here and in what follows small Greek indices $\m,\n$ label external directions which are not extended/doubled, small Latin indices  $a,b,\dots=1,\dots,{\rm dim}\ G$ from  beginning of the alphabet  label generators of Lie algebra $\gg$ of the group manifold $G$, small Latin indices from the middle of the alphabet $m,n,\dots=1,\dots,{\rm dim}\ G$ label coordinates $y^m$ on the group manifold. Functions $\s_m{}^a$ represent components of left-inva\-riant 1-forms on the group manifold and $T_a$ form basis of the corresponding Lie algebra $\gg$. Isometry transformations act on the group manifold from the left as
\begin{equation}
    g \to ug,\quad u\in G.
\end{equation}
Unpacking these notations on may write for the first term in the sigma-model action
\begin{equation}
    \begin{aligned}
        E^{\ha}\wedge * E^{\hb} \h_{\ha\hb}=&\ (g^{-1}dg)^a\wedge *(g^{-1}dg)^bG_{ab} \\
        &+2 (g^{-1}dg)^a \wedge *dx^\m G_{a\m}+ dx^\m dx^\n G_{\m\n},
    \end{aligned}
\end{equation}
where one defines metric components
\begin{equation}
    \begin{aligned}
     G_{\m\n}&=E_\m{}^{\ha}E_\n{}^{\hb}\h_{\ha\hb},\\
     G_{mn}&=\s_m{}^a \s_n{}^b \, G_{ab}= \s_m{}^a \s_n{}^bE_a{}^{\ha}E_b{}^{\hb}\h_{\ha\hb},\\
     G_{m\m}&=\s_m{}^a \, G_{a\m}= \s_m{}^a E_a{}^{\ha}E_\m{}^{\hb}\h_{\ha\hb}.\\
    \end{aligned}
\end{equation}
The 2-form Kalb-Ramond field $B$ is defined as usual as pullback of the corresponding target space-time 2-form field
\begin{equation}
    \begin{aligned}
    B=&\ (g^{-1}dg)^a\wedge (g^{-1}dg)^b B_{ab}\\
    &+2 (g^{-1}dg)^a\wedge dx^\m B_{a\m} + dx^\m \wedge dx^\n B_{\m\n}.
    \end{aligned}
\end{equation}
The fields $G_{ab}, B_{ab}$ are usually referred to as undressed fields as these are free of dependence on group coordinates $y^m$, which has all been left in the 1-forms $\s^a$.

The procedure of NATD of the sigma-model action then proceeds with replacing $(g^{-1}dg)^a \to A^a$ and adding a Lagrange multiplier $\ty_a F^a$. Performing integration over $\ty_a$ one recovers the initial action, while integrating over $A^a$ one turns to a dual action, that now has no dependence on $y^m$ since the 1-forms $\s^a$ no longer present. Instead, a dependence on $\ty_a$ enters the dual background originating from
\begin{equation}
    F^a=2dA^a-f_{bc}{}^a A^b\wedge A^c,
\end{equation}
where $f_{ab}{}^c$ encode structure constants of $\gg$.

This procedure can be summarised nicely by presenting a generalisation of Buscher rules, explicitly providing dual background fields. For that one defines a matrix
\begin{equation}
    \label{Ndef}
    N_{ab}=G_{ab}-B_{ab}+\ty_c f_{ab}{}^c,
\end{equation}
alongwith its inverse $N^{ac}N_{cb}=\d^a{}_b$. The transformation rules are then written as follows
\begin{equation}
    \label{nabuscher}
    \begin{aligned}
        G'_{\m\n} & = G_{\m\n} - (G-B)_{a (\m } N^{ab} (G+B)_{\n) b} \\
        G'_{\m a} & =  \frac{1}{2} (G-B)_{\m b} N^{b a} -  \frac{1}{2} N^{a b}(G-B)_{b \m} \\
        G'_{ab} & =  N^{(ab)} \\
        B'_{\m\n} & = B_{\m\n} + (G-B)_{a [\m } N^{ab} (G+B)_{\n] b} \\
        B'_{\m a} & =  -\frac{1}{2} (G-B)_{\m b} N^{b a} -  \frac{1}{2} N^{a b}(G-B)_{b \m} \\
        B'_{ab} & = - N^{[ab]} 
    \end{aligned}
\end{equation}
These have been shown  in \cite{Catal-Ozer:2019hxw} to be upliftable to the double field theory formalism where the transformation of the fields becomes an O$(d,d)$ matrix with $d={\rm dim}\, G$ as expected. 

\subsection{Double field theory perspective}

Non-abelian T-duality transformation of a 10-dimensional (group manifold) background as described above is known to be equivalent to a sequence of a B-shift and T-duality transformations, equivalently, O(d,d) reflections \cite{Sakatani:2019jgu}. The procedure can be generalised to coset spaces as well, where one chooses $d$ Killing vectors in a $d$-dimensional space and makes basically the same steps. Crucial is that the symmetry group acts without isotropy. In the present text we focus at the case of group manifolds to illustrate the procedure and to make further analysis of its restrictions simpler. Given the results of \cite{Sakatani:2019jgu}, generalisation to coset spaces must be straightforward. 

\begin{figure}[ht]
	\centering
	\begin{tikzpicture}
	\node at (0,3) (BG10) {BG$_{10}$};
	\node at (3,3) (BG10def) {$\widetilde{\rm BG}_{10}$};
	\node at (6,3) (BG10p) {BG$'_{10}$};

	\node at (1.5,3) [above=0.1cm] {\footnotesize $B$-shift};
	\node at (4.5,3) [above=0.0cm] {$T_1,T_2,\dots $};
	\node at (3,4.5) [above=0cm,] {\footnotesize NATD};
    
    \node at (0,2.5) [] {\footnotesize IIA};
    \node at (3,2.5) [] {\footnotesize IIA};	
    \node at (6,2.5) [] {\footnotesize IIB};

	\draw[->, thick, blue] (BG10) edge[out=45, in=135] (BG10p);
	\draw[->, thick, red] (BG10) -- (BG10def);
    \draw[->, thick, blue!80!black] (BG10def) -- (BG10p);
	\end{tikzpicture}
	\caption{
	Relationship between backgrounds upon non-abelian T-duality. $T_i$ denotes usual T-duality along $i$'th direction.
	}
	\label{fig:natd}
\end{figure}
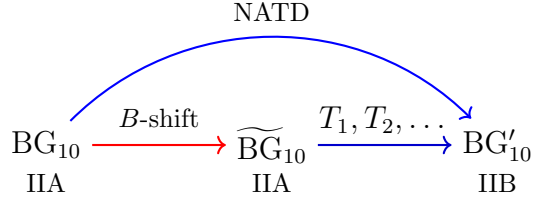

One starts with noticing, that to generalise the NATD transformation rules written in the form \eqref{nabuscher} to 11d backgrounds these can be conveniently rewritten in terms of O(d,d) rotation of a DFT background. Following \cite{Sakatani:2019jgu} the algorithm is as follows
\begin{itemize}
    \item undress background fields;
    \item perform B-shift $B_{ab} \to B_{ab}+ \ty_c f_{ab}{}^c$, with $\ty_a$ understood as coordinates dual to $y^m$.
    \item perform formal abelian T-dualities along all directions of the group manifold to turn $\tilde{y}_a$ into geometric coordinates.
\end{itemize}
Schematically the procedure is depicted on Fig.\ref{fig:natd}.

For further reference and to setup notations let us consider the procedure in more details. The first step splits coordinate dependence to external coordinates $x^\m$ and group manifold coordinates $y^m$ hidden in 1-forms $\s_m{}^a$
\begin{equation}
    \begin{aligned}
        G_{mn}(x,y)&=\s_m{}^a(y)\s_n{}^b(y)G_{ab}(x), \\ B_{mn}(x,y)&=\s_m{}^a(y)\s_n{}^b(y)B_{ab}(x).
    \end{aligned}
\end{equation}
Further B-shift introduces additional dependence dual coordinates $\ty_a$ that is not obvious to check against section constraint. However, one notices that the dependence on $y^m$ is of very restricted form hidden in the 1-forms $\s^a$. For this reason working with undressed fields allows to overcome this issue. Below we show that on explicit examples for both DFT and ExFT, while here we will try to develop some intuition allowing to work with such transformation. 

One starts with an abelian T-duality transformation in the DFT formalism that corresponds to replacing $x^m$ by $\tx_m$, or better to say, to switching their roles as geometric and non-geometric coordinates. Most transparently this is seen when considering doubled pseudo-interval\footnote{Note, that this expression is neither invariant under generalised coordinate transformations, nor represents any distance measurement. Rather, pseudo-interval serves as a convenient form of encoding roles of components of generalised metric.}
\begin{equation}
    \begin{aligned}
        ds^2&=\mH_{MN}d\XX^M d\XX^N\\
            &=\mH_{mn}dy^m dy^n +2\mH_m{}^ndx^m d\ty_n + \mH^{mn}d\ty_m d \ty_n.
    \end{aligned}
\end{equation}
Here and in what follows capital Latin indices $M,N,\dots$ label directions of the extended space and in case of DFT run $1,\dots,2{\rm dim}\ G$. Assigning to $y^m$ and $\ty_m$ the roles of geometric and dual coordinates respectively, one thus fixes $\mH^{mn}=g^{mn}$. To perform T-duality  transformation one keeps the pseudo-interval the same, switching instead roles of coordinates. Say $y^1$ now becomes dual, while $\ty_1$ becomes geometric. This implies, that $\mH_{11}=\tilde{g}^{11}$ is now component of the transformed metric. This procedure has been employed to generate exotic brane solutions and to unify them into a single DFT/ExFT solution in \cite{Berkeley:2014nza,Berman:2014jsa,Berman:2014hna,Bakhmatov:2016kfn}.

For the case in question it is tempting to  writes instead
\begin{equation}
    \label{Hforms}
    \begin{aligned}
        ds^2&=\mH_{MN}d\XX^M d\XX^N\\
            &=\mH_{ab}\s^a \s^b +2\mH_a{}^b \s^a d\ty_b + \mH^{ab}d\ty_a d \ty_b,
    \end{aligned}
\end{equation}
where dependence on $y^m$ has been recollected into the 1-forms $\s^a=\s^a{}_mdx^m$. For now, the dual coordinates are still represented by exact forms $d\ty_a$.

The procedure described guarantees, that one ends up with a solution of supergravity equations of motion if started with a solution. Indeed, let us start for simplicity with a background, that depends purely on group manifold coordinates, i.e. $G_{ab}$ = const, $B_{ab}$ = const. Hence, one may encode the invariant 1-forms in a generalised vielbein $E_M{}^A$ and the constant metric and B-field in the constant ``flat'' generalised metric $\mH_{AB}$
\begin{equation}
    \begin{aligned}
    \mH_{MN}&=E_M{}^AE_N{}^B \mH_{AB}, && 
    \mH_{AB}=
    \begin{bmatrix}
        G_{ab}-B_{ae}B^e{}_b & B_a{}^d \\
        B_b{}^c & G^{cd}
    \end{bmatrix},
    \end{aligned}
\end{equation}
where $G^{ab}$ is simply the inverse of $G_{ab}$. Turning to flux formulation of DFT \cite{Geissbuhler:2013uka} one calculates generalised flux of the generalised vielbein $\mF_{ABC}=3 E_{[A}{}^M E_B{}^N \dt_M E_{C]N}$ and finds the only non-vanishing component $\mF_{ab}{}^c=f_{ab}{}^c$. As expected, this is proportional to the structure constants of the algebra of the 1-forms $\s^a$. Given the initial background is a solution of supergravity equations of motion, such generalised flux and the undressed ``flat'' generalised metric $\mH_{AB}$ are supposed to satisfy equations of motion of DFT in flux formulation.

For the NATD procedure one starts with the undressed fields packed into the ``flat'' generalised metric $\mH_{AB}$ and first preforms a B-shift, that can be encoded as
\begin{equation}
    \begin{aligned}
        \mH_{AB}'&=O_A{}^C(\ty)O_B{}^D(\ty) \mH_{CD}, && O_A{}^B =
        \begin{bmatrix}
            \d_a{}^b & \ty_e f_{ab}{}^e\\
            0        & \d_c{}^d
        \end{bmatrix}.
    \end{aligned}
\end{equation}
Instead of T-dualising all coordinates and checking equations of motion of supergravity, double field theory allows to check $\mH_{AB}'$ explicitly, which is much simpler due to linear dependence on the dual coordinates $\ty_a$. Indeed, in the above expression the matrix $O_A{}^B$ can be understood as a generalised vielbein, and the corresponding generalised flux precisely has the same non-vanishing components $\mF'_{ab}{}^c=f_{ab}{}^c$ as that for $E_M{}^A$. Since the ``flat'' generalised metric $\mH_{AB}$ is the same, the background encoded in $\mH_{AB}'$ satisfies equations of motion of double field theory. Note, that the B-shift is arranged is such a way as to generate a background with the same generalised flux $\mF'_{ABC}=\mF_{ABC}$, which however depends only on dual coordinates. Finally, performing T-dualities along all of $\ty_a$'s one obtains a supergravity solution, since replacing $\ty \leftrightarrow y$ with the corresponding transformation of fields (Buscher rules) is a symmetry of DFT. It is worth mentioning, that after T-dualities the flux components change and one finds non-vanishing $\mF_c{}^{ab}$ components, since T-duality along each direction replaces ${}_a \leftrightarrow {}^a$ \cite{Shelton:2005cf}.

Before turning to an illustrating example, one observes, that the last step where all directions of the group manifold get T-dualised is crucial for ending up with a supergravity solution. It is clear, that one is always able to perform the necessary set of T-dualities to turn all $\ty_a$ into geometric coordinates. Picture however gets more complicated in the case of non-abelian U-dualities and such a set may not exist. We discuss this important point in more details in Section \ref{sec:alg}.

\subsection{Bianchi II example}

As an explicit illustration of the above procedure, consider  the standard examples of Bianchi II cosmological space-time embedded into 10 dimensions. The metric is can be chosen to be 
\begin{equation}
    \begin{aligned}
    ds^2=&\ ds_6^2-a_1{}^2 a_2{}^2 a_3{}^2 (dx^0)^2\\
    &+a_1{}^2(\s^1)^2+a_2{}^2(\s^2)^2+a_3{}^2 (\s^3)^2,
    \end{aligned}
\end{equation}
where the 1-forms $\s^a$ and the functions $a_a$ read
\begin{equation}
    \begin{aligned}
        & \s^1 = dy^1 - y^3 dy^2,&& a_1{}^2 = \fr{p_1}{\cosh(p_2 x^0)} \\
        & \s^2 = dy^2, , && a_2{}^2 = \cosh(p_1 x^0) e^{p_2 x^0},\\
        & \s^3 = dy^3, && a_3{}^2 = \cosh(p_1 x^0) e^{p_3 x^0},          
    \end{aligned}
\end{equation}
and the constants are constrained by $p_2 p_3 = p_1^2$. In what follows we set $p_a = 1$ to avoid the dilaton. Note, that the 1-forms only depend on the coordinates $y^1,y^2,y^3$ on the group manifold generated by the Heisenberg-Weyl algebra
\begin{equation}
    \begin{aligned}
        d\s^a = f_{b c}{}^a \s^b\wedge \s^c, && f_{23}{}^1=1.
    \end{aligned}
\end{equation}
The undressed metric is then
\begin{equation}
    ||g_{ab}||=\mbox{diag}\left[-a_1{}^2 a_2{}^2 a_3{}^2,a_1{}^2,a_2{}^2,a_2{}^2,1,\dots,1\right]  .  
\end{equation}
Since the time direction $x^0$ is not dualised and the metric does not have mixed $g_{0a}$ components, it is enough to focus only at the block $1,2,3$ and consider O$(3,3)$ double field theory. The corresponding generalised metric is simply given by
\begin{equation}
    \mH_{AB} = 
        \begin{bmatrix}
            g_{ab}-B_{ae}g^{ef}B_{fb} && B_{a}{}^d\\ 
            B_{b}{}^{c} && g^{cd},
        \end{bmatrix}
\end{equation}
where capital Latin indices from beginning of the alphabet $A,B,\dots$ represents doubled indices of undressed fields. B-shift is performed by the matrix
\begin{equation}
    O^A{}_B=
        \begin{bmatrix}
            \d^a{}_b && 0\\
            \D B_{cb} && \d^d{}_c
        \end{bmatrix}
\end{equation}
with $\D B_{ab}=\ty_c f_{ab}{}^c$ whose only non-vanishing components are
\begin{equation}
    \D B_{23} = \ty_1.
\end{equation}
Next one is supposed to perform abelian T-dualities along all directions $\ty_a$. T-dualising along all three directions renders all $x^a$ non-geometric as well as the corresponding forms, and one reproduces the well known result for the dual background \cite{Hong:2018tlp}
\begin{equation}
\label{natdB}
    \begin{aligned}
        ds'{}^2=&\ ds^6-a_1{}^2a_2{}^2a_3{}^3(dx^0)^2,\\
        &+\frac{1}{a_1{}^2} (d\ty_1)^2+\frac{a_3{}^2}{\D^2}(d\ty_2)^2+\frac{a_2{}^2}{\D^2}(d\ty_3)^2\\
        B'=&-\frac{\ty_1}{\D^2}d\ty_2 \wedge d\ty_3,\\
        \D^2=&\ a_2{}^2 a_3{}^2+\ty_1^2.
    \end{aligned}
\end{equation}
Note that $\tx_a$ are now proper physical coordinates. The dilaton is recovered from the invariant dilaton
\begin{equation}
    e^{-2\f}\sqrt{g}=e^{-2d}=e^{-2\f'}\sqrt{g'},
\end{equation}
where $g=\det||g_{ab}||$ is determinant of the undressed metric.

\section{Non-abelian U-duality in SL(5) ExFT}
\label{sec:naud}

Let us now try to generalise the above algorithm of NATD to the case of exceptional field theory. As the very first example one may take SL$(5)$ exceptional field theory, that is a $7+10$-dimensional field theory, local coordinate transformations include U-dualities of $D=7$ maximal supergravity \cite{Baguet:2015xha,Musaev:2015ces} (for a review on exceptional field theories see \cite{Hohm:2019bba, Musaev:2019zcr, Berman:2020tqn}). Space-time is split into 7 external directions labelled by coordinates $x^\m$, 4 internal coordinates $y^m$ and 6 dual coordinates $\ty_{mn}=-\ty_{nm}$ corresponding to winding modes of the M2-brane. The latter form the 10-dimensional extended space parametrised by $\XX^{MN}=-\XX^{NM}$, on which generalised Lie derivative is defined. Closure of the algebra of generalised Lie derivatives imposes section condition on all fields and their combinations, that schematically can be written as
\begin{equation}
    \label{section}
    \e^{MNKLP}\dt_{MN}\bullet \dt_{KL}\bullet  =0.
\end{equation}
Field content of the theory can be written in irreps of the duality group SL(5) as follows
\begin{equation}
    \begin{aligned}
        &g_{\m\n}, && A_\m{}^{[MN]}, && m_{(MN)}, && B_{\m\n M},
    \end{aligned}
\end{equation}
where the generalised metric $m_{MN}$ parametrises the coset space ${\rm SL(5)}/{\rm SO(5)}$. Explicit paramterisation in terms of supergravity fields depends on the choice of frame, that is dictated by a choice of solution of the section constraint. Despite the straightforward minimal choice $\dt_{MN}=0$ giving $D=7$ ungauged maximal supergravity, one finds two distinct maximal solutions of the section constraint. These correspond to breaking of the set $\XX^{MN}$ labelling the $\bf 10$ of SL(5) w.r.t. subgroups GL(5) and ${\rm GL(3)}\times {\rm SL(2)}$. The former turns SL(5) ExFT into 11d supergravity, while the latter gives Type IIB supergravity in S-duality covariant formulation.

For the purpose of this letter, we are interested in relations between fields in 11D and IIB frames recovered from explicit parametrisations of the generalised metric $m_{MN}$ and relation of the external metric $g_{\m\n}$ to the $7\times 7$ block of the full 11/10-dimensional metric. One starts with 11-dimensional metric written in the 7+4-split
\begin{equation}
    \begin{aligned}
    ds_{11}^2&=\hgr_{\m\n}dx^\m dx^\n+ \hhr_{mn} dy^m dy^n\\
            &=\hgr_{\m\n}dx^\m dx^\n+ \hhr_{ab} \s^a \s^b.
    \end{aligned}
\end{equation}
Then one has for the ExFT fields $g_{\m\n}$ and $m_{AB}$ 
\begin{equation}
    \begin{aligned}
    g_{\m\n}&=\hhr^{\fr15}\hgr_{\m\n},\\
    m_{AB}&=\hhr^{\fr1{10}}
        \begin{bmatrix}
        \hhr^{-\fr12}\hhr_{ab} & V_{a}\\
        V_b & \hhr^{\fr12}(1+V^2)
        \end{bmatrix},
    \end{aligned}
\end{equation}
where $\hhr= \det ||\hhr_{ab}||$ and the vector $V^a$ encodes internal components of the 3-form field $V^a=\hhr^{-\fr12}\e^{abcd}C_{bcd}$. Note that $\det m_{AB}=1$ and is parametrised by undressed fields.

To recover fields of Type IIB supergravity one switched to the parametrisation corresponding to the GL(3)$\times$SL(2) solution of the section constraint, keeping the ExFT fields the same. For that one has
\begin{equation}
\label{iib}
    \begin{aligned}
        g_{\m\n}&=e^{-\fr45 d}\tgr_{\m\n},\\
        m_{AB}&=e^{-\fr25 d}
            \begin{bmatrix}
                \thr^{\fr12}\thr^{ab}+e^{-2d}\M^{ij}V_i{}^a V_j{}^b & V_i{}^a \\
                V_j{}^b & e^{2d} \M_{ij}
            \end{bmatrix},\\
        e^{-2d}&=e^{-2\f} \thr^{\fr12}.
    \end{aligned}
\end{equation}
Here $d$ is the invariant dilaton of double field theory, $
\thr_{ab}$ is the 3-dimensional block of the full 10-dimensional metric and the matrix $\mM_{ij}$ encodes the degrees of freedom of the axion-dilaton
\begin{equation}
\label{iib2}
    ||\M_{ij}||=
        \begin{bmatrix}
        1 & C_0\\
        C_0 & e^{-2\f} + C_0^2
        \end{bmatrix}.
\end{equation}
The pair of vectors $V_i{}^a$ encode internal parts of the NS-NS Kalb-Ramond 2-form $B_{ab}$ and RR field $C_{ab}$ as
\begin{equation}
    \begin{aligned}
        V_i{}^a=
            \thr^{-1/2}\e^{abc}\begin{bmatrix}
                C_{bc}    \\
                B_{bc}
            \end{bmatrix},
    \end{aligned}
\end{equation}
where $\e^{abc}$ is the Levi-Civita symbol $\e^{123}=1$ It is important to notice, that the parametrisation used here differs from that of \cite{Blair:2020ndg} by rescaling of the metric and 2-form fields by certain power of $e^\f$. More precisely, the parametrisation of \cite{Blair:2020ndg} provides formulation of IIB supergravity explicitly covariant under the SL(2) duality symmetry, that is reflected in the fact, that all dependence on the dilaton is hidden inside the SL(2)/SO(2) matrix. In contrast, the parametrisation given above provides fields T-dual to the IIA fields, that can be obtained from the standard 11D parametrisation. For the purpose of this paper, the latter is more convenient.

Now, following the analogy between DFT and ExFT extended spaces one proposes the following non-abelian U-duality scheme for 11D backgrounds
\begin{enumerate}
    \item undress the metric and the C-field $g_{mn}=\s_m{}^a \s_n{}^b g_{ab}$, $C_{mnk}=\s_m{}^a \s_n{}^b \s_k{}^c C_{abc}$ and compose generalised metric $\mH_{AB}$ from the undressed fields
    \item perform C-shift of the undressed fields by $\D C_{abc}=-3\ty_{d[a}f_{bc]}{}^d$, where $\ty_{ab}$ are the would be dual coordinates
    \item perform a U-duality transformation that turns  $\ty_{ab}$ into geometric coordinates and $\s^a$ into dual 1-form. Equivalently: embed gl(4) in a different way.
\end{enumerate}
As in the case of NATD transformations represented as a B-shift plus T-dualities the above procedure guarantees to always give a solution to equations of motion of 11D supergravity, however with additional restriction to unimodular groups, i.e. $f_{ab}{}^{b} =0$. Origin of the latter condition will become clear momentarily.

The proof is a straightforward repetition of that for double field theory. Starting with a solution to 11D equations of motion one undresses all fields and composes a ``flat'' generalised metric $m_{AB}$ of exceptional field theory. The corresponding generalised vielbein $E_M{}^A$ contains only components of the left-invariant 1-forms $\s_m{}^a$ and hence the only non-vanishing generalised flux components are those, proportional to the structure constants $f_{ab}{}^c$. Using the usual expressions for fluxes of the SL(5) ExFT as in \cite{Blair:2014zba,Berman:2012uy} one finds for the components of the $\bf 10,\, 15$ and $\bf \overline{40}$
\begin{equation}
    \begin{aligned}
        & S_{a5}= 2 f_{ba}{}^b, && \q_{a5}=\fr12 f_{ba}{}^b, && \tilde{T}_{ab5}{}^c=-f_{ab}{}^c-\fr23\d_{[a}{}^c f_{b]d}{}^d.
    \end{aligned}
\end{equation}
As before, C-shift turning $m_{AB}$ to $m'_{AB}$ can be understood as a generalised vielbein with the following generalised fluxes 
\begin{equation}
    \begin{aligned}
        & S'_{a5}= 4 f_{ba}{}^b, && \q'_{a5}= -f_{ba}{}^b, && \tilde{T}'_{ab5}{}^c=-f_{ab}{}^c-\fr23\d_{[a}{}^c f_{b]d}{}^d.
    \end{aligned}
\end{equation}
Hence, generalised fluxes of such constructed background with flat metric and the C-field linearly depending on the dual coordinates $y_{ab}$ are the same as that of the initial background only when $f_{ab}{}^b=0$. Up to this condition, in terms of generalised fluxes and the undressed generalised metric $m_{AB}$ nothing has been changed, and the background $m'_{AB}$ solves equations of motion of  ExFT. Obviously, U-dualising all winding coordinates one ends up with a solution of 11D equations of motion. Arguments for the case, where undressed fields depend on external coordinates $m_{AB}=m_{AB}(x)$ go along the same lines, with however more involved equations of motion in the flux formulation of ExFT. 

\begin{figure}[ht]
	\centering
	\begin{tikzpicture}
	\node at (0,6) (BG11) {BG$_{11}$};
	\node at (3,6) (BG11def) {$\widetilde{\rm BG}_{11}$};
	\node at (0,4) (BG10) {BG$_{10}$};
	\node at (3,4) (BG10def) {$\widetilde{\rm BG}_{10}$};
	\node at (6,4) (BG10p) {BG$'_{10}$};

	\node at (1.5,6) [above=0.1cm] {\footnotesize C-shift};
	\node at (1.5,4) [above=0.1cm] {\footnotesize B-shift};
	\node at (4.3,4) [above=0.0cm] {$T_0,T_1,T_2$};
    
    \node at (0,5) [right] {\footnotesize IIA};
    \node at (3,5) [right] {\footnotesize IIA};	
    \node at (4.5,5) [above=0cm, rotate=-35] {\footnotesize IIB section};
    
    \node at (-1.5,6.05) [right] {\footnotesize 11D:};
    \node at (-1.5,4.05) [right] {\footnotesize 10D:};	
	
	\draw[->, thick, red] (BG11) -- (BG11def);
	\draw[->, thick] (BG11def) -- (BG10p);
	\draw[->, thick] (BG11) -- (BG10);
	\draw[->, thick] (BG11def) -- (BG10def);
	\draw[->, thick, red] (BG10) -- (BG10def);
    \draw[->, thick, blue!80!black] (BG10def) -- (BG10p);

	\end{tikzpicture}
	\caption{
	Relationship between backgrounds with spectator fields upon the non-abelian U-duality procedure. Here taking a IIB section represents an uplift of three T-dualities with further reduction to 10 dimensions. In this case the bottom line represents the usual non-abelian T-duality. 
	}
	\label{fig:naud}
\end{figure}
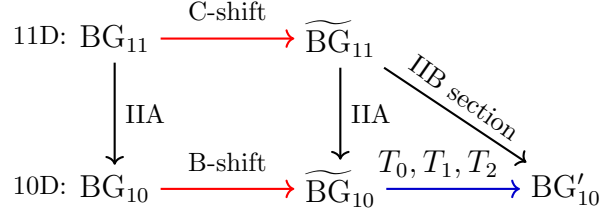

Although here we restrict ourselves to the case of SL(5) ExFT for simplicity, the first two steps of the procedure have straightforward generalisation to higher U-duality group simply by including more winding coordinates. In contrast, the last step appears to be much more restricted for the SL(5) theory than for theories with more winding directions. As we show below, at least for group manifolds full dualisation of all four coordinates is possible only when at least one of the coordinates is an abelian isometry, i.e. corresponds to a spectator field. We will conclude that the described procedure for the SL(5) ExFT is always an uplift of an NATD transformation. Similar observations based on the construction of exceptional Drinfled algebras for the SL(5) theory have been made in \cite{Blair:2020ndg}.   Schematically, this is illustrated on Fig. \ref{fig:naud}.

\section{Algebraic perspective}
\label{sec:alg}

\subsection{T-duality}

Non-abelian T-duality as a particular case of Poisson-Lie T-duality is based on the notion of Drinfeld double Lie algebra, that is basically a Manin triple $(\gg,\tgg,\h)$, with $\gg$ and $\tgg$ being Lie algebras with bases $\{T_a\}=\bas\,\gg$ and $\{\tT^a\}=\bas\,\tgg$, and $\h$ a non-degenerate invariant symmetric bilinear form on $\gg\oplus \tgg$. Defining $\{ T_A \}=\{ T_a,\tT^a\}=\bas\,\gg\oplus\tgg$ the commutation relations read
\begin{equation}
 [T_A, T_B]=\mF_{AB}{}^C T_C,
\end{equation}
where the only non-vanishing structure constants are $\mF_{ab}{}^c=:f_{ab}{}^c$ and $\mF^{ab}{}_c =: \tf^{ab}{}_c$. Non-vanishing components of the $ad$-invariant symmetric form  $\h(T_A, T_B) =\h_{AB} $ are then $\h^a{}_b=\d^a{}_b=\h_b{}^a$. 

Geometrically such defined Drinfeld double can be realised by choosing a maximally isotropic subalgebra say $\gg$ to be a ``physical'' subalgebra. Group element $g=\exp x^a T_a$ of the corresponding Lie group $G$ defined by generators of the physical subalgebra will define left-invariant 1-forms $\s=g^{-1}dg$ on the group manifold. In this setup a non-abelian  T-duality corresponds to transfer the rope of the physical subalgebra to the dual algebra $\tgg$ and constructing space-time 1-forms from group element $\tilde{g}=\exp x_a \tilde{T}^a$. Note, that here $x_a$ is a physical coordinate and not further T-duality is required. More generally a Poisson-Lie T-duality is a constant transformation of the generators $T_A$ preserving the bilinear form $\h$, i.e. an O$(d,d)$ rotation $T_A'=C_A{}^B T_B$, under which a given Drinfeld double is invariant. Split of the Drinfeld double into the set of physical and dual coordinates breaks O$(d,d)$ to the physical GL$(d)$ upon which the vector representation decomposes as
\begin{equation}
    2 {\bf d} \longrightarrow \bf{d} \oplus \bar{\bf d}.
\end{equation}
The $\bf d$ will be chosen to correspond to the physical subalgebra, while the alternative choice corresponds to taking $\bf \bar{d}$. The corresponding physical algebra $\mathfrak{gl}(d)$ will be embedded into $\mathfrak{o}(d,d)$ differently and switching between these two is represented by the external automorphism of $\mathfrak{o}(d,d)$ corresponding to deleting one of the two spinorial roots. 

More transparently this is seen when looking at  generalised vielbeins $E_A{}^M$ with the inverse defining the generalised metric $\mH_{MN}=E_M{}^AE_N{}^B \mH_{AB}$. These can be explicitly constructed in the component form in terms of the left-invariant 1-forms and a $B$- or $\b$-field. For us important is that the generalised veilbein $E_A{}^M$ realises the same double Drinfeld algebra w.r.t. to the generalised Lie derivative 
 (see \cite{Sakatani:2019jgu} for more details)
\begin{equation}
    [E_A,E_B]=\mF_{AB}{}^C E_C.
\end{equation}
Hence, a transformation $C_A{}^B$ can be understood as acting on the algebraic indices of the generalised vielbein. An NATD transformation is equivalent to a B-shift and a series of  O$(d,d)$ reflections along all $d$ directions, performed on the undressed generalised vielbein $E_M{}^{A}$. From the DFT point of view the latter are necessary to turn all $\tilde{x}_a$ to geometric coordinates, while on the Drinfeld double language this replaces all generators $T_a$ by $\tT^a$. The set of $d$ T-dualities interchanging $\bf d$ and $\bf \bar{d}$ (normal and winding coordinates) can be equivalently understood and choosing a different embedding of the maximal GL$(d)$ subgroup, such that $\bf \bar{d}$ becomes its fundamental and $\bf d$ its co-fundamental representations. This indeed corresponds to deleting one of the two sinorial roots of the Dynkin diagram of O$(d,d)$.

One notices, that according to the B-shift+T-dualities procedure, one has to replace all winding coordinates by their geometric partners, which can be done in a unique way for O$(d,d)$ theory (for groups-manifolds that are not a product of Lie groups). This seems to be in tension with the Poisson-Lie T-plurality picture, where a given Drinfeld double can be decomposed into a set of more than two Manin triples \cite{Snobl:2002kq}. Backgrounds corresponding to such Manin triples generate the same Drinfeld double and hence are indistinguishable from the point of view of the two-dimensional sigma model. Examples of such backgrounds can be found in \cite{Hlavaty:2019pze}. In the O$(d,d)$ language Poisson-Lie T-plurality corresponds to performing a rotation by an O$(d,d)$ matrix $C_A{}^B$, preserving the Drinfeld double, which in particular can be a set of $d$ reflections \cite{Sakatani:2019jgu}. This latter case is precisely the transformation, that turns all winding coordinates into geometric ones. Hence, in all other cases one would expect backgrounds, which do not solve equations of motion of normal supergravity due to remaining dependence on winding coordinates. Indeed, as has been shown on explicit examples in \cite{Hlavaty:2019pze} such procedure in particular gives solutions of generalised supergravity equations. More generally, one always ends up with a DFT background. Simply speakin, equivalent $\mathfrak{gl}(d)$ embeddings into $\mathfrak{o}(d,d)$ can be obtained from a given one by O$(d,d)$ rotations and by the external automorphism of the algebra. Only the latter turns the fundamental of a given embedding of $\gl(d)$ into the antifundamental of the dual embedding. Crucial here is that no weight belongs to both these representations, which is apparent for the $\mathfrak{o}(d,d)$ algebra but is not always true for symmetry algebras of exceptional field theories.

To conclude, one starts with an irrep $\mR_1$ of the abelian T(U)-duality group in which extended coordinates transform. Upon an embedding of the geometric GL$(d)$ subgroup this decomposes into $\mR_1 \to {\bf d} \oplus \dots $ where $\bf d$ corresponds to geometric coordinates and ellipses denotes irreps under which winding coordinates transform. Now, one considers a different embedding of the geometric GL$(d)$ such that $\mR_1 \to {\bf d'} \oplus \dots $, where $\bf d'$ is the fundamental of GL$(d)$ none of whose weights inside $\mR_1$ coincide with any of the weights of $\bf d$. Let us provide more details for U-duality groups SL(5), where this cannot be done, and SO(5,5), that can be shown to allow 11-dimensional NAUD.

\subsection{U-duality and exceptional Drinfeld algebras}

\begin{figure}[ht]
\centering
\begin{tikzpicture}


\draw (1,0) node (1) {$\m_1$}; 
\draw (3,0) node (2) {$\m_2$};
\draw (5,0) node (3) {$\m_3$};
\draw (7,0) node (4) {$\m_4$}; 
\draw (9,0) node (5) {$\m_5$};

\draw[->, blue] (4) edge (5);
\draw[->] (3) edge (4);
\draw[->, red] (1) edge (2);
\draw[->, \bgreen] (2) edge (3) ;

\draw[dashed] (2,-0.5) -- (2,0.5);
\draw (2,-0.8) node[rotate=-90] (sc2) {\Leftscissors \footnotesize{2}};

\draw[dashed] (8,-0.5) -- (8,0.5);
\draw (8,-0.8) node[rotate=-90] (sc1) {\Leftscissors \footnotesize{1}};

\end{tikzpicture}
\caption{Weight diagram of the fundamental  $\bf 5$ of $\sl(5)$ with the highest weights represented by $\m_{1}$. The action of different roots is denoted by different colours and the direction of arrows shows the lowering of the weight. Depending on the chosen deletion of a simple root one gets two different decompositions $\bf 5 \to 4_0 + 1_{-4}$ under $\sl(5) \hookleftarrow \gl(4)$}
\label{fig:5}
\end{figure}
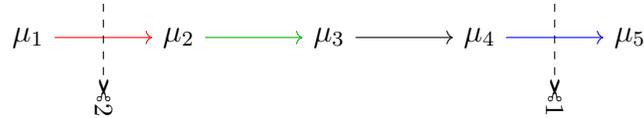

We start with the set of the simple roots of the Lie algebra $\mathfrak{sl}(5)$ in the canonical $\w$-basis of fundamental weights are given by the following
\begin{equation}
\begin{aligned}
\a_{12}&=(2,-1,0,0),\\
\a_{23}&=(-1,2,-1,0),\\
\a_{34}&=(0,-1,2,-1),\\
\a_{45}&=(0,0,-1,2),
\end{aligned}
\end{equation}
where the labelling of the roots will become clear momentarily. The remaining positive roots are
\begin{equation}
\begin{aligned}
&\a_{13}=\a_{12}+\a_{23}, \qquad \a_{14}=\a_{12}+\a_{23}+\a_{34}, \\
&\a_{24}=\a_{23}+\a_{34}, \qquad \a_{25}=\a_{23}+\a_{34}+\a_{45},\\
&\a_{35}=\a_{34}+\a_{45}, \qquad \a_{15}=\a_{12}+\a_{23}+\a_{34}+\a_{45}.
\end{aligned}
\end{equation}
In addition, one has the same  number of negative roots and four Cartan generators. Weight diagram of the fundamental representation $\bf 5$ of $\sl(5)$ is depicted on Fig. \ref{fig:5}, where $\m_1,\dots,\m_5$ denote basis vectors. Notations for the simple root of the algebra are chosen in such a way that, say the root $\a_{12}$ sends the weight vector $\m_1$ to $\m_2$. Or, equivalently, the exponent $\exp(\w \a_{12})$ acts by SL(2) rotations on the plane $(\m_1,\m_2)$.

\begin{figure}[ht]
\centering
\begin{tikzpicture}


\draw (1,0) node[circle, minimum height=0.5cm, red, draw, thick] (a12) {$ $};
\draw (3,0) node[circle, minimum height=0.5cm, \bgreen,  thick, draw] (a23) {$ $};
\draw (5,0) node[circle, minimum height=0.5cm, draw, thick] (a34) {$ $};

\draw (7,0) node[circle, minimum height=0.5cm, blue, draw, thick] (a45) {$ $};

\draw (1,0.2) node[above=0.1] (12) {$\a_{12}$}; 
\draw (3,0.2) node[above=0.1] (23) {$\a_{23}$}; 
\draw (5,0.2) node[above=0.1] (34) {$\a_{34}$}; 
\draw (7,0.2) node[above=0.1] (45) {$\a_{45}$}; 

\draw (1.6,-0.6) node[rotate=-45] (sc2) {\Leftscissors \footnotesize{2}};
\draw[dashed] (0.6,0.4) -- (1.4,-0.4);

\draw (7.6,-0.6) node[rotate=-45] (sc1) {\Leftscissors \footnotesize{1}};
\draw[dashed] (6.6,0.4) -- (7.4,-0.4);

\draw[-] (a12) edge (a23);
\draw[-] (a23) edge (a34);
\draw[-] (a34) edge (a45);

\end{tikzpicture}
\caption{Dynkin diagram of $\sl(5)$ with simple roots coloured differently for further convenience. Depending on the two possible ways to delete one root keeping three connected, depicted by \Leftscissors {\footnotesize 1} and \Leftscissors {\footnotesize 2}, one obtains two embeddings of the $\gl(4)$ subalgebra related by external automorphism.}
\label{fig:sl5roots}
\end{figure}
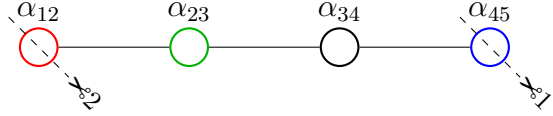

From the Dynkin diagram of $\sl(5)$ on Fig. \ref{fig:sl5roots} one finds two embeddings of the subalgebra $\gl(4)$, corresponding to deleting the root $\a_{12}$ or the root $\a_{45}$. In matrix representation this corresponds to embedding a $4 \times 4$ matrix as an upper left or lower right block. As is shown in Fig. \ref{fig:5} depending on the chosen deletion of a root, one ends up with different decompositions of the fundamental $\bf 5 \to \bf 4 \oplus 1$. It is important to note, that the weights $\m_2,\m_3,\m_4$ belong to a $\bf 4$ for both of the decompositions, while one of $\m_1,\m_5$ becomes a singlet.

\begin{figure}[ht]
\centering
\begin{tikzpicture}

\draw (1,6) node (12) {$\XX^{12}$}; 
\draw (3,6) node (13) {$\XX^{13}$}; 
\draw (5,6) node (14) {$\XX^{14}$}; 
\draw (7,6) node (15){$\XX^{15}$}; 
\draw (3,4) node (23) {$\XX^{23}$};
\draw (5,4) node (24){$\XX^{24}$}; 
\draw (7,4) node (25){$\XX^{25}$};
\draw (5,2) node (34){$\XX^{34}$}; 
\draw (7,2) node (35){$\XX^{35}$}; 
\draw (7,0) node (45){$\XX^{45}$};

\draw[->] (12) edge[\bgreen] (13) (13) edge (14) (14) edge[blue] (15) (13) edge[red] (23) (14) edge[red] (24) (15) edge[red] (25) (23) edge (24) (24) edge[blue] (25)
(24) edge[\bgreen] (34) (25) edge[\bgreen] (35) (34) edge[blue] (35) (35) edge (45);

\draw[dashed] (0.3,5) -- (8,5);
\draw (0,5) node[rotate=0] (sc2) {{\footnotesize{2}}\Rightscissors};

\draw[dashed] (6,0) -- (6,7);
\draw (6,-0.3) node[rotate=90] (sc1) {{\footnotesize{1}}\Rightscissors};

\end{tikzpicture}
\caption{Weight diagram of the $\bf 10$ of $\mathfrak{sl}(5)$ with two possible embeddings of the $\gl(4)$ subalgebra.}
\label{fig:10}
\end{figure}
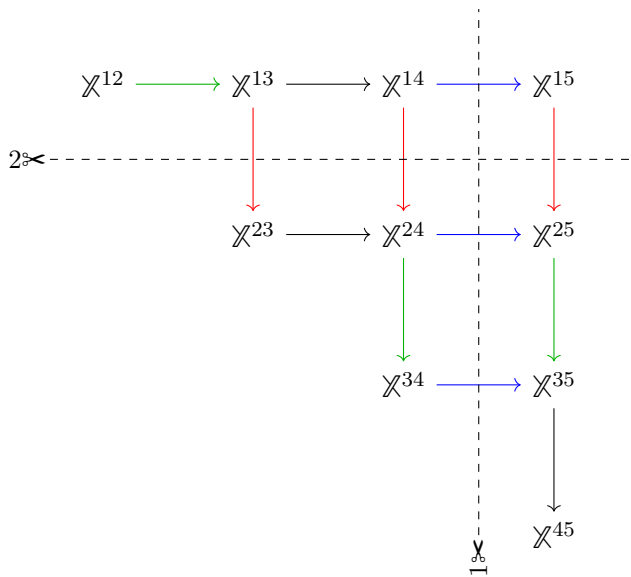

Following the analogy with the NATD one is interested in embeddings of the physical $\gl(4)$  subalgebra related by the external automorphism. In particular for the SL(5) theory we are interested in decomposing the $\bf 10$ of $\sl(5)$ upon two embeddings of $\gl(4)$, that are shown in Fig. \ref{fig:10}. Consider first the decomposition corresponding to deleting the root $\a_{45}$ (cutting blue arrows). In this case weight vectors $\XX^{5a}$ with $a=1,\dots,4$ belong to the $\bf 4$ of $\gl(4)$ while the rest $\XX^{ab}$ belong to the $\bf 6$. In the ExFT language, the former get identified with geometric coordinates, while the latter represent winding modes.

Now, according to the procedure of NAUD described above, one need to find such a different embedding of $\gl(4)$, that all weights contributed to the irrep governing geometric coordinates of the first embedding belong to that governing winding modes. Explicitly, all weights from the old $\bf 4$ must belong to the new $\bf 6$, which is impossible, according to Fig. \ref{fig:10}. 

Indeed, suppose one starts with four left-invariant 1-forms $\s^a$, that depend on four coordinates on the (unimodular) group manifold $x^1,x^2,x^3,x^4$. Next one constructs a background with flat metric and C-field given by $C_{abc}=-3 \ty_{d[a}f_{bc]}{}^d$ with $\ty_{ab}=1/2 \e_{abcd}\XX^{cd}$ being coordinates along winding directions. This has been shown to solve equations of motion of ExFT, however to end up with an ordinary supergravity solution one has introduce such turn all $\ty_{ab}$ into geometric coordinates such that all $\XX^{5a}$ become non-geometric. From Fig. \ref{fig:10} one concludes that the automorphism acts by interchanging indices $1 \longleftrightarrow 5$ upon which the directions $\XX^{25},\XX^{35},\XX^{45}$ become non-geometric since belong to the new $\bf 6$, while $\XX^{15}$ belongs to the new $\bf 4$ and hence must be thought of as a geometric direction. According to the speculative discussion around \eqref{Hforms} these directions correspond to 1-forms, rather than coordinates and hence the forms $\s^{2},\s^{3},\s^{4}$ must be thought of as ``non-geometric'' while $\s^1$ as a ``geometric''. It is suggestive to understand a (non-)geometric 1-form components as those, which depend on (non-)geometric coordinates. Unless $d\s^1 = 0$, one ends up with a contradiction, when the same set of coordinates on which $\s^{a}$ depend should be understood as non-geometric and as geometric at the same time.

The above conditions can be fulfilled when $T_1$ commutes trivially with the rest three generators. In this case $d\s^1=0$ and it can be chosen to depend say on $x^1$ one which the other 1-forms $\s^\a$ with $\a=2,3,4$ do not depend. Indeed, otherwise $d\s^\a$ would give $\s^1$ on the RHS generating non-vanishing $f_{1 a}{}^{\a}$. One concludes, that the described procedure applied to a 4-dimensional group manifold provides a solution of supergravity equations of motion only when at least one spectator field is included. This is in consistency with observations made in \cite{Blair:2020ndg}. Another option would be to generalise the notion of T-plurality to the case of non-abelian U-duality. From the DFT point of view T-plurality generates backgrounds with dependence on dual coordinates, which in particular cases solve generalised supergravity equations. However, no generalised supergravity extension to 11 dimensions is known, and moreover this is widely accepted to not exist.

More strict and rigorous formulation of these points is required and it is tempting to believe that this can be achieved in the formalism of DFT${}_{\rm WZW}$ \cite{Hassler:2017yza,Blumenhagen:2014gva,Hassler:2015pea}.More detailed investigation of such formulations is reserved for future work.

\definecolor{amethyst}{rgb}{0.6, 0.4, 0.8}

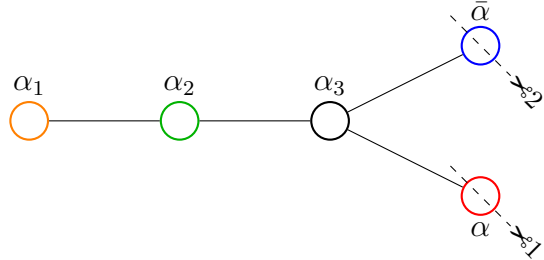
\begin{figure}[ht]
\centering
\begin{tikzpicture}


\draw (1,0) node[circle, minimum height=0.5cm, orange, draw, thick] (a12) {$ $};
\draw (3,0) node[circle, minimum height=0.5cm, \bgreen,  thick, draw] (a23) {$ $};
\draw (5,0) node[circle, minimum height=0.5cm, draw, thick] (a34) {$ $};
\draw (7,-1) node[circle, minimum height=0.5cm, red, draw, thick] (a) {$ $};
\draw (7,1) node[circle, minimum height=0.5cm, blue, draw, thick] (ab) {$ $};

\draw (1,0.2) node[above=0.1] (12) {$\a_{1}$}; 
\draw (3,0.2) node[above=0.1] (23) {$\a_{2}$}; 
\draw (5,0.2) node[above=0.1] (34) {$\a_{3}$}; 
\draw (7,-1.2) node[below=0.1] (45) {$\a$}; 
\draw (7,1.2) node[above=0.1] (45) {$\bar{\a}$};

\draw (7.6,0.4) node[rotate=-45] (sc2) {\Leftscissors \footnotesize{2}};
\draw[dashed] (6.6,1.4) -- (7.4,0.6);

\draw (7.6,-1.6) node[rotate=-45] (sc1) {\Leftscissors \footnotesize{1}};
\draw[dashed] (6.6,-0.6) -- (7.4,-1.4);

\draw[-] (a12) edge (a23);
\draw[-] (a23) edge (a34);
\draw[-] (a34) edge (a);
\draw[-] (a34) edge (ab);

\end{tikzpicture}
\caption{Dynkin diagram of $\so(5,5)$ with simple roots coloured differently for further convenience. Depending on the two possible ways to delete one root keeping three connected, depicted by \Leftscissors {\footnotesize 1} and \Leftscissors {\footnotesize 2}, one obtains two embeddings of the $\gl(5)$ subalgebra. }
\label{fig:so55roots}
\end{figure}

Consider now more fruitful case of five dimensions and U-duality algebra $\so(5,5)$. Its Dynkin diagrams with two possible deletions of simple roots giving $\gl(5)$ is depicted on Fig. \ref{fig:so55roots}. This has three simple roots generating vector representation, antisymmetric tensor of second and third rang representations and two spinorial representations. 

Embeddings of $\gl(5)$ are recovered by deleting one of the two spinorial roots: $\a$ or $\bar{\a}$. Here we focus on the $\bf 16$ of $\so(5,5)$ under which coordinates of the SO(5,5) ExFT transform and which governs transformations of generators of the SO(5,5) exceptional Drinfeld algebra $(T_a, T^{ab}, T)$ with $a,b=1,\dots,5$. Weight diagram for the spinorial representation $\XX^M$ with $M=1,\dots,16$ is given on Fig. \ref{fig:16}
\begin{figure}[ht]
\centering
\begin{tikzpicture}

\draw (1,6) node (1) {$\XX^{1}$}; 
\draw (3,6) node (2) {$\XX^{2}$}; 
\draw (5,6) node (3) {$\XX^{3}$}; 
\draw (7,6) node (4){$\XX^{4}$}; 
\draw (9,6) node (5){$\XX^{5}$}; 

\draw (5,4) node (6) {$\XX^{6}$};
\draw (7,4) node (7){$\XX^{7}$}; 
\draw (9,4) node (8){$\XX^{8}$};

\draw (7,2) node (9){$\XX^{9}$}; 
\draw (9,2) node (10){$\XX^{10}$}; 
\draw (11,2) node (11){$\XX^{11}$};

\draw (7,0) node (12) {$\XX^{12}$};
\draw (9,0) node (13){$\XX^{13}$}; 
\draw (11,0) node (14){$\XX^{14}$};
\draw (13,0) node (15){$\XX^{15}$}; 
\draw (15,0) node (16){$\XX^{16}$};

\draw[->] (1) edge[blue] (2) (2) -- (3) (3) edge[\bgreen] (4) (4) edge[orange] (5) (3) edge[red] (6) (4) edge[red] (7) (5) edge[red] (8) (6) edge[\bgreen] (7) (7) edge[orange] (8)  
(7) -- (9) (8) -- (10) (9) edge[orange] (10) (10) edge[\bgreen] (11) (9) edge[blue] (12) (10) edge[blue] (13) (11) edge[blue] (14)
(12) edge[orange] (13) (13) edge[\bgreen] (14) (14) -- (15) (15) edge[red] (16)

;

\draw[dashed] (4.3,5) -- (10,5);
\draw (4,5) node[rotate=0] (sc1) {{\footnotesize{1}}\Rightscissors};

\draw[dashed] (14,-0.7) -- (14,1);
\draw (14,-1) node[rotate=90] (sc1) {{\footnotesize{1}}\Rightscissors};

\draw[dashed] (6.3,1) -- (12,1);
\draw (6,1) node[rotate=0] (sc2) {{\footnotesize{2}}\Rightscissors};

\draw[dashed] (2,5.3) -- (2,7);
\draw (2,5) node[rotate=90] (sc2) {{\footnotesize{2}}\Rightscissors};

\end{tikzpicture}
\caption{Weight diagram of the $\bf 16$ of $\so(5,5)$ with two possible embeddings of the $\gl(5)$ subalgebra.}
\label{fig:16}
\end{figure}

Now, from the diagram it is clear that upon the first embedding the geometric coordinates (equivalently, ``physical'' generators of the SO(5,5) EDA) correspond to the weights $(\XX^1,\dots,\XX^5)$, while the rest correspond to winding modes. Upon the second embedding the ``physical'' subaglebra of EDA is spanned by generators corresponding to the weights $(\XX^{12},\dots, \XX^{15})$. One notices, that the two sets of physical coordinates do not intersect and one is able to perform such an SO(5,5) transformation as to shift all 1-forms $\s^a$ into the non-geometric set. Equivalently, this demonstrates existence of two possible choices of the ``physical'' subalgebra inside exceptional Drinfeld algebra with SO(5,5) symmetry, which do not conflict.

\section{Discussion}

In this letter a generalisation of the non-abelian T-duality Buscher rules for 10D supergravity backgrounds to 11D backgrounds has been proposed. For that one starts with the representation of the conventional NATD as a B-shift of  undressed  generalised metric linearly proportional to dual coordinates $\D B_{ab}=f_{ab}{}^c \ty_c$ with further abelian T-dualities along all directions to turn all $\ty_a$ into geometric coordinates. Naturally this translates into a procedure that starts with C-shift of the generalised metric of exceptional field theory $\D C_{abc}= -3 \ty_{d[a}f_{bc]}{}^d$, that produces a field configuration depending on dual coordinates. To end up with a solution of supergravity equations one either performs a formal conventional U-duality transformation that turns dual coordinates into geometric, or chooses appropriate IIB section. These procedures can be understood as construction of a background with flat metric and gauge fields linearly depending on dual coordinates such that it has precisely the same generalised fluxes as the initial one. Such defined background is then guaranteed to solve equations of motion of double (exceptional) field theory and hence of the usual supergravity upon T(U)-duality of all winding directions. For the NAUD case the procedure has been checked to work only for backgrounds with unimodular symmetry group, i.e. group manifolds with $f_{ab}{}^b=0$. For NATD such backgrounds would generate non-vanishing trombone gauging, which in general would require generalised supergravity framework. Similar observation can be made in the ExFT case. Indeed, the tension between generalised flux components of the SL(5) theory due to terms containing $f_{ab}{}^b$ can be removed by passing a linear dependence on dual coordinates to the field $\f$ proportional to determinant of the external metric. At the level of fluxes this has many similarities with the dilaton $d$ of DFT, whose linear dependence on dual coordinates gives rise to generalised supergravity. However, it is widely accepted generalised supergravity does not exist in 11 dimensions based on the observation that no Weyl symmetry to break presents for the membrane. Observations made in the present work suggest further investigation of possible relaxation of this statement.

From the algebraic point of view the set of T-dualities along all directions is equivalent to replacing generators $T_a$ by their duals $\tilde{T}^a$ in the double Drinfeld algebra (Manin triple). For supergravity backgrounds that means that one embeds the ``physical'' $\gl(d)$ in two different ways: such that $T_a$ or $\tilde{T}^a$ transform in the fundamental $\bf d$ of $\gl(d)$. This corresponds to external automorphism of the $\mathfrak{o}(d,d)$ algebra replacing deletion one of the roots on either ends by the deletion of the root on the opposite end. This observation and the requirement for U-dualisation of all dual coordinates suggests to understand NAUD transformation as a switch between two ``physical'' algebras $\gl(d)$ by external automorphism of the corresponding exceptional symmetry algebra. We show, that for the algebra $\sl(5)$ such procedure can generate solutions of  the conventional supergravity only when a spectator field presents, which is in consistency with observation made in \cite{Blair:2020ndg}. Investigating the  example of the algebra $\so(5,5)$ one concludes that larger U-duality symmetry groups allow such non-abelian U-dualisation and a solution of equations of motion of 11-dimensional supergravity can be constructed. Investigation of explicit examples based on the SO(5,5) and E${}_6$ exceptional Drinfeld algebra is reserved to future work.

One becomes naturally interested in generalisation of the obtained results to exceptional field theories to general manifolds with isometries along the line of \cite{Sakatani:2019jgu,Sakatani:2019zrs,Malek:2019xrf}. In this case symmetries manifest themselves in the algebra of Killing vectors, which can be used to organise an tri-vector shift, in contrast to the 3-form shift in the present paper \cite{Bakhmatov:2019dow, Bakhmatov:2020kul}. This provides tri-vector deformations of 11-dimensional backgrounds, which in certain cases follow the same scheme as in Figure \ref{fig:naud}. E.g. one considers tri-vector deformation of Minkowski space-time, which in the IIB frame is again a Minkowski space-time, while solves equations of motion of generalised supergravity in the IIA frame \cite{Bakhmatov:2019dow}. More detailed analysis of relations between deformations and non-abelian dualities is required.

\section{Acknowledgements}

The author thanks vivid discussions with I. Bakhmatov, K. Gubarev, E. Malek and N. Sadik Deger that motivated this project. The author thanks Yuho Sakatani for useful comments and suggestions. This work was supported by the Foundation for the Advancement of Theoretical Physics and Mathematics ``BASIS'' and by Russian Ministry of education and science (Project 5-100). In part the work was funded by the Russian Government program of competitive growth of Kazan Federal University.  

\appendix

\bibliography{bib.bib}
\bibliographystyle{utphys.bst}

\end{document}